# Automating Analysis of Construction Workers' Viewing Patterns for Personalized Safety Training and Management

Idris Jeelani[a], Kevin Han[a] and Alex Albert[a]

[a] Department of Civil, Construction & Environmental Engineering, North Carolina State University, Raleigh, USA
E-mail: idrisj@ncsu.edu, kevin_han@ncsu.edu, alex_albert@ncsu.edu

**Abstract –**
Unrecognized hazards increase the likelihood of workplace fatalities and injuries substantially. However, recent research has demonstrated that a large proportion of hazards remain unrecognized in dynamic construction environments. Recent studies have suggested a strong correlation between viewing patterns of workers and their hazard recognition performance. Hence, it is important to study and analyze the viewing patterns of workers to gain a better understanding of their hazard recognition performance. The objective of this exploratory research is to explore hazard recognition as a visual search process to identifying various visual search factors that affect the process of hazard recognition. Further, the study also proposes a framework to develop a vision based tool capable of recording and analyzing viewing patterns of construction workers and generate feedback for personalized training and proactive safety management.

**Keywords –**
Hazard Recognition, Construction Safety, Computer vision, Eye tracking

## 1 Introduction

Every day about 6.5 million workers work on more than 200,000 construction sites across the United States [1]. However, with 900 fatal and over 200,000 non-fatal injuries a year [1], construction workers are more likely to be injured than their counterparts in other industries. Over the years, researchers have attempted to understand the underlying precursors of accidents in construction [2,3]. Among others, hazard recognition has received significant attention [4–6]. Recognizing hazards is one of the first steps in effective safety management; however, studies across the world have shown that 30% to 50% of hazards remain unrecognized in construction environments [6–10].

To improve hazard recognition skill of workers, several training programs have been developed and implemented within the construction industry. However, research has shown that construction workers fail to recognize a significant portion of hazards despite having received training [11,12]. Research in human factors and visual search [13] has revealed several contributing factors that influence person's detection and recognition performance while they scan their environment for hazards. For example. a recent study [14] argues that one of the reasons for unrecognized hazards is that workers often pay attention to the primary task alone and ignore the surrounding areas. The same study also suggests that workers often selectively pay attention to a particular type of hazards while ignoring others types, which result in poor hazard recognition levels.

To recognize hazards in their workplaces, workers need to first detect the hazardous object or condition and identify it correctly, to associate a level of risk to it. This is essentially a visual search process and the outcome of any such process is affected by the viewing pattern and attention distribution of the person conducting the search [15]. Therefore, it is important to design training interventions that aim to improve workers' visual search and enable them to detect hazard better. The first step towards developing such training interventions is to understand what constitutes a good visual search.

Hence, there is a need to explore hazard recognition from a visual search perspective to identify and evaluate various metrics that define the viewing patterns of workers and affect their hazard recognition performance. In order to gain an intrusive understanding of workers' viewing behavior, it is important to study their viewing patterns while they engage in their regular work in real construction environments. However, examining the viewing patterns of workers manually would not only require a lot of time and repetitive effort but will also be prone to human errors causing inaccurate assessment and inefficient feedback. Therefore, the objective of this study is 1) to identify the visual search metrics that affect hazard recognition and 2) automate the capture and analysis of visual search data collected from workers in real construction sites.

More specifically, the study provides a framework for a computer-vision based system that can be used to generate personalized feedback for workers and safety managers for training purposes and generate data that will potentially assist in proactive safety management.

## 2 Background

### 2.1 Visual Search

Visual search is a task of scanning the environment to find particular visual stimuli or features (the targets) among other visual stimuli or features (the distractors) [16]. Hazard recognition can essentially be viewed as a "multiple target visual search process" where hazards are the targets while other objects and features in the environment serve as the distractors.

The visual search is influenced by several factors, such as age differences [17], professional experience [18], the similarity in targets and distractors [19], and the type of search being conducted [16]. Therefore, studying visual search allows us to examine attentional abilities, cognition and perception.

Eye-tracking provides an objective measure of stimuli that received attention during visual search activities by analyzing eye movement data and has found application in aviation, medicine, transportation, and education [20,21]. The Eye tracking device records the corneal reflection of infrared lighting to track pupil position, mapping the subject's focus of attention on his field of view (gaze) [22].

### 2.2 Computer Vision in Construction

Computer vision seeks to enable computer systems to automate the tasks that the human visual system do [23–28]. Computer vision techniques include image reconstruction, object detection and tracking, feature extraction, pose estimation, image restoration etc. One of the most popular applications is perhaps the face detection in our cell phone cameras.

Computer vision is widely used in robotics[23], medicine [24], surveillance [25], transportation [26] and others. In construction, computer vision is gaining popularity due to its applications in progress monitoring[29–31], productivity analysis[27,32], structural health monitoring, 3D reconstruction [33,34] and automated documentation [35,36]. Although computer vision has not been leveraged to its full potential, it has several applications in replacing or augmenting the conventional field-based safety monitoring tools [37].

## 3 Objectives

The objectives of this research were:
1) to identify the visual search metrics that affect hazard recognition performance of workers and
2) to provide a framework for developing an automated vision-based system that records the gaze behavior of workers in a construction site and computes the visual search metrics identified above.

Objective 1 helps us understand the process of hazard recognition from a visual search perspective and Objective 2 enables us to collect and analyze that information automatically for a large-scale personalized safety monitoring and training. Achieving these two goals will enable interventions and proactive safety management while maintaining a high manager-to-worker ratio that is economically inevitable.

## 4 Research Method

The study was completed in two phases (fig 1). Phase 1 consisted of identifying the visual search metrics that affect hazard recognition performance. Phase 2 consisted of developing a computer vision based tool to capture and analyze eye-tracking data in order to compute the identified metrics automatically and on a large scale.'

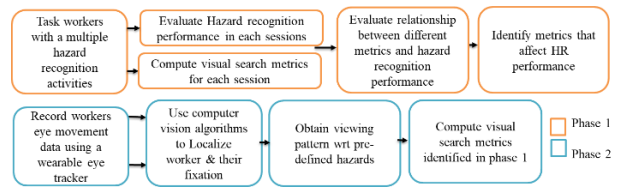

Fig 1. Method Overview

### 4.1 Phase 1

This phase was completed in two steps as detailed below

#### 4.1.1 Data Collection

Twenty-four construction workers representing diverse speciality trades were selected. The experience of workers varied from 2 years to 28 years (Mean: 10.05) and their ages ranged from 19 years to 51 Years (Mean: 32.5). After gathering worker demographic information, the participants were placed 30 cm from a computer screen with the centre of the screen in line with participant's eyes. An Eye tracking device (Eyetech VT-2) with the sampling rate of 60 Hz placed at the top of the screen was calibrated for each worker.

The workers were then tasked with identifying hazards in 12 pre-selected construction case images that were randomly selected from a set of 24 case images captured from real projects. The images were displayed on a computer screen with a resolution of 1920 X 1080. As the workers identified hazards verbally from each case image, the researchers catalogued the information and their eye movements were captured and recorded by the eye tracker.

The hazards in each case image were pre-identified by an expert panel of three safety professionals with the cumulative experience of 62 years. The hazard recognition performance of workers was measured in terms of hazard recognition index:

$$HRI = \frac{Hazards\ identified\ by\ subject}{Total\ Preidentified\ Hazards\ in\ that\ image}$$

The eye movement data captured during the experiment was used to calculate following visual search metrics:

*Search Duration (SD)*

Search duration is the time (in seconds) for which the participant views a scene before terminating the search. The search duration consists of fixations (when participants focus on a particular point) and saccades (rapid eye movements between fixations)

*Fixation Count (FC)*

The fixation count is the number of discrete points in a visual field, where participant fixates or focuses their attention. The number of fixations is indicative of the importance or "noticeability" of objects/ areas to a user [38].

*Fixation Time (FT)*

Fixation time is the time spent by each participant in fixating on different objects or areas. It is calculated by taking the sum of durations of all fixations. Fixation time reflects the amount of attention paid by the participant to each photograph.

$$T = \sum_{i=1}^{n}(E(f_i) - S(f_i))$$

Where E and S are the end time and start time for i$^{th}$ fixation ($f_i$)

*Mean Fixation Duration (MFD)*

The mean fixation duration (MFD), also known as average fixation duration [39], is the sum of the durations of all fixations divided by the number of fixations[40]. It is computed using equation [37]

$$MFD = \frac{FT}{FC}$$

This metric indicates the time spent by the participants focusing on individual objects. Longer fixation durations indicate that subjects spent more time and effort in evaluating and extracting information from the stimulus [41]. Longer fixations are also indicative of more visual effort and "substantial increase in demand for attentiveness [42]

*Ratio of On-Target: All-target Fixation Time (ROAFT)*

Each hazard in the case images was defined as the area of Interest (AOI), which is an area that fully encloses the hazard. The Ratio of "On-Target" to "all target" fixation time or ROAFT [43] is the sum of durations of fixation on AOI, divided by the total duration of all fixations for the entire case image (Area of Glance or AOG). It is computed using equation [43]

$$ROAFT = \frac{\sum_{i=1}^{n}(E(f_i) - S(f_i))\ in\ AOI}{\sum_{j=1}^{n}(E(f_j) - S(f_j))\ in\ AOG}$$

Where E and S are the end time and start time for i$^{th}$ fixation ($f_i$) in AOI and j$^{th}$ Fixation in AOG in numerator and denominator respectively.

This metric is a good indicator of the amount of visual attention devoted to hazards relative to the total time spent on the case image.

*Fixation Rate (FR)*

Fixation rate is the ratio of the number of fixation in AOI to the total number of fixation in the entire case image (that is Area of Glance or AOG).

$$FR = \frac{f_n}{f_N}$$

Where $f_n$ = the number of fixations in AOI
$f_N$ = Total number of fixations in AOG

The fixation rate is indicative of search efficiency, with smaller ratio suggesting that participants spend more effort in finding the pertinent objects, thereby indicating lower search efficiency [38].

### 4.1.2 Data analysis

After collecting the data, the correlation analysis was carried out for each worker to measure the correlation between each of the metrics listed above and hazard recognition performance.

For each worker, average HRI index was calculated using equation:

$$AV\_HRI_j = \frac{\sum_{i=1}^{n} HRI_{ji}}{n}$$

Where $AV\_HRI_j$ = average hazard recognition performance for j$^{th}$ worker
$HRI_{ji}$ = Hazard recognition index for jth worker in ith photograph
$n$ = number of photographs = 12

Similarly, for each visual search metric described above, the average value was computed for each worker. Finally, Pearson's correlation coefficients were calculated for each metrics to obtain each of degree of correlation with hazard recognition performance.

### 4.2 Phase 2

Once relevant metrics are identified in phase 1, this phase focused on developing the framework for an automated vision based system that captures eye-tracking data from workers while they engage in their regular work and compute these identified metrics.

The framework of the system can be demonstrated in two sub phases- Pre-processing phase and test phase as shown in the figure 2 below.

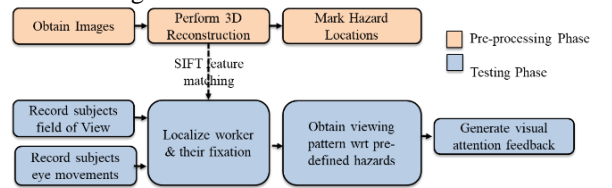

Figure 2. Method overview for Phase 2

### 4.2.1 Pre-processing phase

This phase focuses on building the pre-processing data. It is completed in two steps as detailed below

#### 4.2.1.1 Obtaining initial reference images

Multiple images of a construction site are obtained using a monocular camera. The easiest way to obtain such images is to record a walkthrough video of construction site while following different trajectories. This ensures that objects are captured from different angles and distances. The frames are then extracted from this video, which serve as reference images to be used in subsequent steps.

#### 4.2.1.2 Annotating AOIs

In this step, a panel of safety experts pre-identifies the hazards in the work site by watching the walkthrough video and conducting open discussions. The locations of these AOIs are marked on an initial set of images as AOIs. This can be semi-automated using homography matrices and affine transformation (discussed later in the case study). The hazards are annotated on preprocessing images as rectangular bounding boxes. The bounding boxes are defined by Xmin, Ymin & Xmax, Ymax, which correspond to the upper left corner and lower right corner of the bounding box respectively.

### 4.2.2 Testing Phase

This phase focuses on computing various visual search metrics for individual workers as they move in a real world construction site. This involves localizing the participant and their gaze positions at regular time intervals. This helps in defining participants' attention distribution and in turn compute various visual search metrics. The steps involved in testing phase are as follows:

#### 4.2.2.1 Obtain test images

As participants move around the construction site, a wearable eye tracker equipped with an HD video camera can simultaneously record their gaze position & direction, and their field of view as first-person view (FPV). The image frames are then extracted from the recorded video, which serve as test images to be used in subsequent sections.

#### 4.2.2.2 Localizing workers

In this step, the position of a worker at different intervals is obtained. The test images are appended to the point cloud, developed in pre-processing phase, and each test image is feature matched with the pre-processing images, to find the best match for each test image. Scale Invariant Feature Transform (SIFT) algorithm [44] is used to conduct pairwise matching. These matches (or correspondences) are stored in an array for future use discussed in step 2.4). The best match for a particular test image is the pre-processing image that has maximum matched features with it. The known camera location of the preprocessing image gives us the location of the participant for its best-matched test image. This enables us to track the trajectory of worker throughout the construction site.

#### 4.2.2.3 Obtain gaze location for each frame

The goal of this step is to obtain the fixation point (gaze location) of the subject in each testing image. The wearable eye tracker records, the data stream that contains gaze position of the subject at every 1/100th second. Using the timestamps, the data is synced with the frame rate of the FPV to extract a gaze positions such that they represent each frame of the FPV. Since each frame represents a test image, we have a gaze position for every test image.

#### 4.2.2.4 Transforming gaze positions using Homography matrices

After the 4.2.2.3, we now have a fixation point (gaze position) for each test image and from step 4.2.2.2; we have a corresponding preprocessing image for each test image. Hence, the objective of this step is to transform the fixation point location from test images to the matched preprocessing image. After this, the system checks whether the gaze location is within any of the AOIs or not. If it is, we count the fixation for that AOI and measure the duration of fixation using the timestamps.

To transform the location of fixation point from testing image to its corresponding pre-processing image, we compute the homography matrix for each match [45] using the correspondences obtained in step 4.2.2.2. The homography (H) matrix is calculated using the following equation 1 [45]

$$\boldsymbol{x'_i = Hx_i} \qquad\qquad Eq.\ (1)$$

Where $x_i = [x_i\ y_i\ w_i]^T$ and $x'_i = [x'_i\ y'_i\ w'_i]^T$ are the point correspondences that are obtained through pairwise SIFT feature matching.

#### 4.2.2.5 Obtaining viewing pattern with respect to pre-identified hazards

The following condition is checked for test image with x' y' being the transformed gaze position calculated above

$$X` > X_{min}\ AND\ x` < X_{max}\ AND\ y` > Y_{min}\ AND\ y` < Y_{max}$$

Where $X_{min}$, $Y_{min}$ and $X_{max}$, $Y_{max}$ are the lower left and upper right corners of bounding box defining the particular AOI.

If the condition is met, it indicates that the fixation point is within the AOI. In this case, the fixation is counted for the particular AOI, if the condition remains true for at least 240ms (average fixation duration in a visual search process). The duration of fixation is obtained by calculating the difference between the timestamp when the condition was met and the last timestamp before the check fails.

#### 4.2.2.6 Analyzing the data and computing metrics

Finally, the visual attention distribution for each subject is generated. Specifically, the output provides information about the areas subjects paid attention to and areas they did not. This information is used to calculate various visual search metrics identified in phase 1.

## 5 Results

### 5.1 Phase 1

The results of correlation analysis are shown below.

Table 1. Results of Phase 1

| Metric | Pearson's Correlation Coef | p-value | N |
| --- | --- | --- | --- |
| SD | 0.563** | 0.005 | 23 |
| FT | 0.635** | 0.001 | 23 |
| FC | 0.649** | 0.001 | 23 |
| MFD | 0.393* | 0.064 | 23 |
| ROAFT | -0.093 | 0.673 | 23 |
| FR | -0.132 | 0.548 | 23 |

** Significant at $\alpha = 0.05$, * Significant at $\alpha = 0.07$.
The results indicate that hazard recognition performance is strongly and positively correlated with search duration, fixation counts, and fixation time. MFD shows a weak correlation, which is also significant at $\alpha = 0.07$. This implies that searching for longer durations, focusing visual attention on more number of objects and areas, and fixating for longer durations can improve hazard recognition performance.

The results do not provide enough evidence to suggest that a relationship between hazard recognition and remaining two visual search metrics exist (at α<0.05).

### 5.2 Phase 2

#### 5.2.1 Case study Set up

The framework for automated capture and analysis of eye tracking data presented in this study was validated in a live construction site in Raleigh, North Carolina. The objective was to automatically record and analyze attention distribution of a subject while they move around the construction site wearing a wearable eye-tracking device and validate the accuracy of output by comparing it with manually computed results. Five hazards were pre-identified by a panel of safety experts in the portion of the construction site that was to be used for this case study. These hazards are shown in figure 3

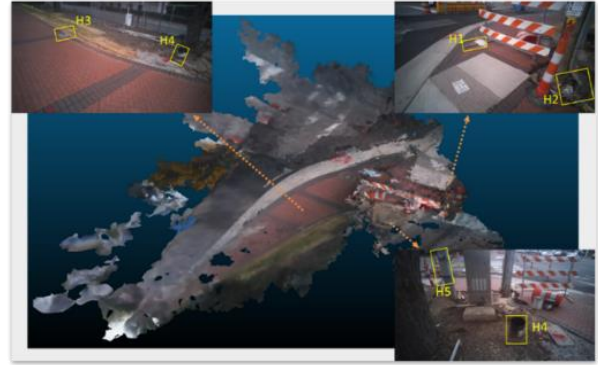

Figure 3. Predefined Hazards. H1: Trip Hazard; H2= Live Electrical Wires; H3: Protruding Rod; H4: Chemical Hazard; H5: Electric Junction Box

The methodology described in 4.2 was followed as explained below.

#### 5.2.1.1 Pre-processing Phase

A walk through video of the selected portion of the site was recorded using Tobii glasses 2 front mounted HD camera with the 10p resolution. A one-minute and 42-second video captured the test area from multiple angles and 2512 frames were extracted that served as pre-processing images.

The features were extracted from these images using SIFT algorithm and these features were used to perform 3D reconstruction of the test area using Structure for Motion or SFM [46]. After this, the pre-identified hazards were annotated on the pre-processing images as AOIs. The AOIs were manually annotated in few images and using the homography between the images, the locations of AOIs were computed automatically on other images using homography matrices.

#### 5.2.1.2 Test Phase:

The subject was tasked to move in the test area while his eye movement data and his field of view were recorded using Tobii glasses 2. It recorded the gaze position and gaze direction, which was used to compute the fixation point location (gaze location) of the subject. This data was recorded at 100Hz and the video was recorded at 25 FPS that captured subject's field of view.

The video recorded by Tobii glasses was exported and 362 frames were extracted from it, which served as test images. The eye tracking data was also exported and the data rows containing the gaze position were extracted. The localization of subject and his gaze position was carried out as explained in the method section (4.2.2.2). Finally, the number of fixations on each AOI and fixation durations were computed as described in step 4.2.2.5. The process was repeated for all test images to obtain the visual attention distribution for the subject while he was in the test area and compute all metrics described in 2.1

#### 5.2.2 Case Study Results

The visual attention distribution for the subject was computed in terms of distribution of dwell times (total fixation duration on an AOI) among five pre-identified AOIs. The other visual search metrics were computed as described in section 4.1. Table 2 & 3 shows the results obtained from the proposed system.

Table 2. Visual Attention distribution of subject

| $DT_1$ (ms) | $DT_2$ (ms) | $DT_3$ (ms) | $DT_4$ (ms) | $DT_5$ (ms) |
|---|---|---|---|---|
| 900 | 235 | 257 | 1148 | 1270 |

Table 3. Visual search metrics computed by system

| SD (ms) | FC | FT (ms) | MFD |
|---|---|---|---|
| 18250 | 33 | 8212.5 | 248.86 |

As shown in Table 2, participant spent 18 seconds in the test area and fixated on 33 different objects/areas. The Dwell times are shown in table 1 DT1 to DT5 give the distribution of attention over different AOS, which indicate that the participant focused mainly on hazard 4 and 5 paid little attention to hazards 2 and 3. The total fixation time (FT) was 8.21 sec, meaning subject spent only 45% of the time in obtaining information from the scene; remaining 7.52 sec or 55% of the time was spent in searching with no acquisition of visual information.

## 6   Validation of Accuracy of Localization

To validate the accuracy of localization of fixation points, the dwell times obtained from the system were compared with the swell times computed manually using fixation overlay video obtained from Tobii's recording software. The fixation count for each AOI was obtained by manually counting the number of sets where fixation was within AOI for six consecutive frames. This was multiplied by the mean fixation duration to get the dwell time for each AOI. The results of validation show an accuracy is 88% on average.

Table 4. Validation Results

| AOI | H1 | H2 | H3 | H4 | H5 |
|---|---|---|---|---|---|
| DT1 System | 900 | 235 | 257 | 1148 | 1270 |
| DT2 Manual | 744 | 248 | 248 | 992 | 992 |
| Variation (ms) | 156 | 13 | 9 | 156 | 278 |
| **Accuracy %** | **83%** | **94%** | **96%** | **86%** | **78%** |

## 7   Discussion & Conclusions

The results of this study advance the theoretical knowledge in the area of construction safety and assist researchers and practitioners develop effective training programs and hazard control strategies. This study represents one of the first research efforts towards understanding hazard recognition from a visual search perspective and evaluating the key parameters that impact hazard search. The contributions and the implications of this study are discussed below.

First, the study identifies and examines various visual search metrics that are closely related to hazard recognition performance. This paves the way for further studies aimed at identifying and studying the effects of various factors that influence workers ability to detect and recognize hazards in construction environments.

Second, the results suggest a strong correlation between the duration of search and the number of hazards recognized by a worker. The current practice does not require workers to spend a specific amount of time in searching for hazards. Instead, workers arbitrarily (or based on experience) examine the work environment, hoping to detect all hazards. Hence, sessions that involve hazard recognition in the field (such as Job Hazard Analysis) should be designed as such to require workers spend a set minimum amount of time to search for hazards in their work environment. Similarly, the training programs can focus on the importance of spending sufficient time in searching hazards as well.

Third, the second phase of the study demonstrates the use of computer vision techniques to automate the process of analyzing eye movement data. It presents a novel framework for developing a system that is capable of analyzing eye-tracking data with dynamic subjects in real environments, which is an active problem in eye tracking research. It helps in automated analyses of viewing behavior and visual attention distribution of construction workers on a large scale. This data can be used to provide focused and personalized process feedback to workers, which will help subject understand their viewing behavior and identify specific deficiencies in their hazard search.

Moreover, the system enables us to collect and analyze visual attention data on a much larger scale. This large data will help researchers better understand the relationship between viewing patterns and hazard recognition performance. The data can also provide insights about common viewing patterns among workers and help safety managers identify hazards that are more likely to remain unrecognized in construction sites.

Finally, the framework would be beneficial to eye tracking researchers in other areas as well, that require analysis of eye tracking data when subjects are moving in real-world environments.

## 8   References

[1] U.D. of Labor, OSHA: Commonly Used Statistics, Osha.gov. (2016). https://www.osha.gov/oshstats/commonstats.html (accessed February 11, 2017).

[2] P. Mitropoulos, T.S. Abdelhamid, G.A. Howell, Systems model of construction accident causation, Journal of Construction Engineering

and Management-Asce. 131 (2005) 816–825. doi:10.1061/(asce)0733-9364(2005)131:7(816).

[3] S. Rajendran, J.A. Gambatese, Development and Initial Validation of Sustainable Construction Safety and Health Rating System, Journal of Construction Engineering and Management-Asce. 135 (2009) 1067–1075. doi:10.1061/(asce)0733-9364(2009)135:10(1067).

[4] R. Sacks, A. Perlman, R. Barak, Construction safety training using immersive virtual reality, Construction Management and Economics. 31 (2013) 1005–1017. doi:10.1080/01446193.2013.828844.

[5] A. Albert, M.R. Hallowell, B.M. Kleiner, Enhancing Construction Hazard Recognition and Communication with Energy-Based Cognitive Mnemonics and Safety Meeting Maturity Model: Multiple Baseline Study, Journal of Construction Engineering and Management. 140 (2014) 4013042. doi:10.1061/(ASCE)CO.1943-7862.0000790.

[6] G. Carter, S.D. Smith, Safety Hazard Identification on Construction Projects, Journal of Construction Engineering & Management. 132 (2006) 197–205. doi:10.1061/(ASCE)0733-9364(2006)132:2(197).

[7] S. Bahn, Workplace hazard identification and management: The case of an underground mining operation, Safety Science. 57 (2013). doi:10.1016/j.ssci.2013.01.010.

[8] M. Namian, A. Albert, C.M. Zuluaga, E.J. Jaselskis, Improving Hazard-Recognition Performance and Safety Training Outcomes: Integrating Strategies for Training Transfer, Journal of Construction Engineering and Management. 142 (2016) 4016048. doi:10.1061/(ASCE)CO.1943-7862.0001160.

[9] I. Jeelani, K. Han, A. Albert, Development of Immersive Personalized Training Environment for Construction Workers, in: Congress on Computing in Civil Engineering, Proceedings, 2017.

[10] I. Jeelani, A. Albert, R. Azevedo, E.J. Jaselskis, Development and Testing of a Personalized Hazard-Recognition Training Intervention, Journal of Construction Engineering and Management. 143 (2017). doi:10.1061/(ASCE)CO.1943-7862.0001256.

[11] R.A. Haslam, S.A. Hide, A.G.F. Gibb, D.E. Gyi, T. Pavitt, S. Atkinson, A.R. Duff, Contributing factors in construction accidents, Applied Ergonomics. 36 (2005) 401–415. doi:10.1016/j.apergo.2004.12.002.

[12] A. Perlman, R. Sacks, R. Barak, Hazard recognition and risk perception in construction, Safety Science. 64 (2014) 13–21. doi:10.1016/j.ssci.2013.11.019.

[13] S.R. Mitroff, A.T. Biggs, The Ultra-Rare-Item Effect: Visual Search for Exceedingly Rare Items Is Highly Susceptible to Error, Psychological Science. 25 (2014) 284–289. doi:10.1177/0956797613504221.

[14] I. Jeelani, A. Albert, J.A. Gambatese, Why Do Construction Hazards Remain Unrecognized at the Work Interface?, Journal of Construction Engineering and Management. 143 (2017). doi:10.1061/(ASCE)CO.1943-7862.0001274.

[15] M.J.J. Wang, S.C. Lin, C.G. Drury, Training for strategy in visual search, International Journal of Industrial Ergonomics. 20 (1997) 101–108. doi:10.1016/S0169-8141(96)00043-1.

[16] A.M. Treisman, G. Gelade, A feature-integration theory of attention, Cognitive Psychology. 12 (1980) 97–136. doi:10.1016/0010-0285(80)90005-5.

[17] D.J. Madden, W.L. Whiting, R. Cabeza, S.A. Huettel, Age-related preservation of top-down attentional guidance during visual search, Psychology and Aging. 19 (2004) 304–309. doi:10.1037/0882-7974.19.2.304.

[18] A.T. Biggs, M.S. Cain, K. Clark, E.F. Darling, S.R. Mitroff, Assessing visual search performance differences between Transportation Security Administration Officers and nonprofessional visual searchers, Visual Cognition. 21 (2013) 330–352. doi:10.1080/13506285.2013.790329.

[19] D. Guest, K. Lamberts, The time course of similarity effects in visual search., Journal of Experimental Psychology: Human Perception and Performance. 37 (2011) 1667–1688. doi:10.1037/a0025640.

[20] N.B. Sarter, R.J. Mumaw, C.D. Wickens, Pilots' monitoring strategies and performance on automated flight decks: An empirical study combining behavioral and eye-tracking data, Human Factors. 49 (2007) 347–357. doi:10.1518/001872007x196685.

[21] T. Tien, P.H. Pucher, M.H. Sodergren, K. Sriskandarajah, G.-Z. Yang, A. Darzi, Eye tracking for skills assessment and training: a systematic review, Journal of Surgical Research. 191 (2014) 169–178. doi:10.1016/j.jss.2014.04.032.

[22] A.T. Duchowski, A breadth-first survey of eye-tracking applications, Behavior Research Methods, Instruments, & Computers. 34 (2002) 455–470. doi:10.3758/BF03195475.

[23] E. Bayro-Corrochano, Geometric computing: for wavelet transforms, robot vision, learning, control and action, Springer Publishing Company, Incorporated, 2010.

[24] N. Ayache, Medical computer vision, virtual

reality and robotics, Image and Vision Computing. 13 (1995) 295–313.

[25] G.A. Jones, N. Paragios, C.S. Regazzoni, Video-based surveillance systems: computer vision and distributed processing, Springer Science & Business Media, 2012.

[26] B. Coifman, D. Beymer, P. McLauchlan, J. Malik, A real-time computer vision system for vehicle tracking and traffic surveillance, Transportation Research Part C: Emerging Technologies. 6 (1998) 271–288.

[27] K.K. Han, M. Golparvar-Fard, Potential of big visual data and building information modeling for construction performance analytics: An exploratory study, Automation in Construction. 73 (2017). doi:10.1016/j.autcon.2016.11.004.

[28] K.K. Han, D. Cline, M. Golparvar-Fard, Formalized knowledge of construction sequencing for visual monitoring of work-in-progress via incomplete point clouds and low-LoD 4D BIMs, Advanced Engineering Informatics. 29 (2015) 889–901. doi:10.1016/j.aei.2015.10.006.

[29] M. Golparvar-Fard, F. Peña-Mora, S. Savarese, Automated Progress Monitoring Using Unordered Daily Construction Photographs and IFC-Based Building Information Models, Journal of Computing in Civil Engineering. 29 (2015) 4014025. doi:10.1061/(ASCE)CP.1943-5487.0000205.

[30] K.K. Han, M. Golparvar-Fard, Appearance-based material classification for monitoring of operation-level construction progress using 4D BIM and site photologs, Automation in Construction. 53 (2015) 44–57. doi:10.1016/j.autcon.2015.02.007.

[31] K.K. Han, D. Cline, M. Golparvar-Fard, Formalized knowledge of construction sequencing for visual monitoring of work-in-progress via incomplete point clouds and low-LoD 4D BIMs, Advanced Engineering Informatics. 29 (2015). doi:10.1016/j.aei.2015.10.006.

[32] J.W. Lee, C.W. Cho, K.Y. Shin, E.C. Lee, K.R. Park, 3D gaze tracking method using Purkinje images on eye optical model and pupil, Optics and Lasers in Engineering. 50 (2012) 736–751. doi:10.1016/j.optlaseng.2011.12.001.

[33] K. Asadi, H. Ramshankar, H. Pullagurla, A. Bhandare, S. Shanbhag, P. Mehta, S. Kundu, K. Han, E. Lobaton, T. Wu, Building an Integrated Mobile Robotic System for Real-Time Applications in Construction, arXiv Preprint arXiv:1803.01745. (2018).

[34] K. Asadi, K. Han, Real-Time Image-to-BIM Registration Using Perspective Alignment for Automated Construction Monitoring, in: Construction Research Congress, 2018.

[35] S.M. Shahandashti, S.N. Razavi, L. Soibelman, M. Berges, C.H. Caldas, I. Brilakis, J. Teizer, P.A. Vela, C. Haas, J. Garrett, B. Akinci, Z. Zhu, Data-Fusion Approaches and Applications for Construction Engineering, Journal of Construction Engineering and Management-Asce. 137 (2011) 863–869. doi:10.1061/(asce)co.1943-7862.0000287.

[36] M. Golparvar-Fard, F. Peña-Mora, C. Arboleda, S. Lee, Visualization of Construction Progress Monitoring with 4D Simulation Model Overlaid on Time-Lapsed Photographs, Journal of Computing in Civil Engineering. 23 (2009) 391–404.

[37] J. Idris, H. Kevin, A. Alex, Scaling Personalized Safety Training Using Automated Feedback Generation, Construction Research Congress 2018. (2018). doi:doi:10.1061/9780784481288.020.

[38] A. Poole, L.J. Ball, Eye Tracking in Human-Computer Interaction and Usability Research: Current Status and Future Prospects, Encyclopedia of Human-Computer Interaction. (2005) 211–219. doi:10.4018/978-1-59140-562-7.

[39] D.D. Salvucci, J.H. Goldberg, Identifying fixations and saccades in eye-tracking protocols, in: Proceedings of the Symposium on Eye Tracking Research & Applications - ETRA '00, 2000: pp. 71–78. doi:10.1145/355017.355028.

[40] T. Busjahn, C. Schulte, A. Busjahn, Analysis of code reading to gain more insight in program comprehension, Proceedings of the 11th Koli Calling International Conference on Computing Education Research - Koli Calling '11. (2011) 1–9. doi:10.1145/2094131.2094133.

[41] G. Cepeda Porras, Y.-G. Guéhéneuc, An empirical study on the efficiency of different design pattern representations in UML class diagrams, Empirical Software Engineering. 15 (2010) 493–522. doi:10.1007/s10664-009-9125-9.

[42] R. Kliegl, A. Nuthmann, R. Engbert, Tracking the mind during reading: The influence of past, present, and future words on fixation durations, Journal of Experimental Psychology: General. 135 (2006) 12–35. doi:10.1037/0096-3445.135.1.12.

[43] J.H. Goldberg, X.P. Kotval, Computer interface evaluation using eye movements: Methods and constructs, International Journal of Industrial Ergonomics. 24 (1999) 631–645. doi:10.1016/S0169-8141(98)00068-7.

[44] D. Lowe, Distinctive Image Features from Scale-Invariant Keypoints, International Journal of Computer Vision. 60 (2004) 91–110.

[45] A. Zisserman, R. Hartley, Multiple view geometry in computer vision, Cambridge University Press, 2003.

[46] VisualSFM : A Visual Structure from Motion System, http://ccwu.me/vsfm/(accessed February 11, 2017).

.